\documentclass[]{aastex701}
\usepackage{amsmath}

\begin{document}

\title{Unexpected Local Enhancement of Chromospheric Moreton Waves on the Sun}

\correspondingauthor{P. F. Chen}
\email{chenpf@nju.edu.cn}

\author[orcid=0000-0002-9908-291X]{Y. W. Ni}
\affiliation{School of Astronomy and Space Science, Nanjing University, Nanjing 210023, China}
\affiliation{Key Laboratory of Modern Astronomy and Astrophysics, Ministry of Education, Nanjing University, Nanjing 210023, China}
\email{}  

\author[orcid=0000-0002-4205-5566]{J. H. Guo} 
\affiliation{School of Astronomy and Space Science, Nanjing University, Nanjing 210023, China}
\affiliation{Key Laboratory of Modern Astronomy and Astrophysics, Ministry of Education, Nanjing University, Nanjing 210023, China}  
\affiliation{KU Leuven, Department of Mathematics, Centre for mathematical Plasma Astrophysics, Celestijnenlaan 200B, B-3001 Leuven, Belgium} 
\email{}

\author[0000-0001-7693-4908]{C. Li}  
\affiliation{School of Astronomy and Space Science, Nanjing University, Nanjing 210023, China}  
\affiliation{Key Laboratory of Modern Astronomy and Astrophysics, Ministry of Education, Nanjing University, Nanjing 210023, China}  
\affiliation{Nanjing University, Institute of Science and Technology for Deep Space Exploration, Suzhou Campus, Suzhou 215163, China}  
\email{}  

\author[0000-0003-1594-3610]{Y. K. Wang}  
\affiliation{School of Astronomy and Space Science, Nanjing University, Nanjing 210023, China}  
\affiliation{Key Laboratory of Modern Astronomy and Astrophysics, Ministry of Education, Nanjing University, Nanjing 210023, China}  
\email{}  

\author[0000-0001-9610-0433]{G. Y. Chen}  
\affiliation{School of Astronomy and Space Science, Nanjing University, Nanjing 210023, China}  
\affiliation{Key Laboratory of Modern Astronomy and Astrophysics, Ministry of Education, Nanjing University, Nanjing 210023, China}  
\email{}  

\author[0000-0002-7153-4304]{C. Xia}  
\affiliation{School of Physics and Astronomy, Yunnan University, Kunming 650500, China} 
\email{}  

\author[]{C. Fang}  
\affiliation{School of Astronomy and Space Science, Nanjing University, Nanjing 210023, China}  
\affiliation{Key Laboratory of Modern Astronomy and Astrophysics, Ministry of Education, Nanjing University, Nanjing 210023, China} 
\email{}  

\author[0000-0002-7289-642X]{P. F. Chen}  
\affiliation{School of Astronomy and Space Science, Nanjing University, Nanjing 210023, China}  
\affiliation{Key Laboratory of Modern Astronomy and Astrophysics, Ministry of Education, Nanjing University, Nanjing 210023, China}  
\affiliation{State Key Lab of Lunar and Planetary Sciences, Macau University of Science and Technology, Macau 999078, China}  
\email{chenpf@nju.edu.cn}

\begin{abstract}

Solar eruptions often drive shock waves. When propagating in the solar corona, strong shock waves would leave imprints on the solar surface in the chromosphere, which are called Moreton waves. Moreton waves generally decay as they move away. However, to our knowledge, for the first time we report that a Moreton wave is significantly enhanced midway during propagation, as revealed by the spectroscopic observations from China's first solar satellite, the Chinese H\(\alpha\) Solar Explorer. Combining magnetic measurements from the Solar Dynamics Observatory, we found that the location of the Moreton wave enhancement corresponds to a magnetic null point at an altitude of $\sim$70 Mm in the corona. It is demonstrated that the unexpected local enhancement of the Moreton wave is due to mode conversion around the magnetic null point, i.e., fast-mode waves are converted to slow-mode waves which are trapped locally. 
It is further revealed that, even for ordinary Moreton wave fronts, the H$\alpha$ Doppler redshift velocity is positively correlated with the height of widespread low-altitude magnetic null points, implying that these low-lying null points,  $\sim$1--7 Mm above the solar surface, facilitate mode conversion with different efficiency, thereby accounting for the spatially inhomogeneous Dopplergram of Moreton wave fronts.

\end{abstract}


\section{Introduction}
Above the visible surface of the Sun lies the solar atmosphere, including the photosphere, chromosphere, and the extended corona higher up. The magnetized solar atmosphere is full of waves due to ceaseless convection below the solar surface, ubiquitous jets, and sporadic eruptions such as solar flares and coronal mass ejections (CMEs). Early in the 1960s, it was discovered that along with solar flares, a wave front propagates away from the flare site with a speed of 500--2000 km s$^{-1}$, which was later called Moreton-Ramsey wave or Moreton wave \citep{Moreton1960, Mor1960}. The leading wave front is manifested as dark in the red wing of the H$\alpha$ spectral line and bright in the blue wing and line center \citep{Dodson1968}. Such characteristics imply that the solar chromosphere, where the spectral line H$\alpha$ ($\lambda$=6562.8~\AA\ ) forms, moves downward at the wave fronts so as to produce Doppler redshifts \citep{Chen2016}. Based on spectral observations, the downward velocity of the chromosphere at wave fronts is around 4--10 km s$^{-1}$ \citep{Svestka1976, Cabezas2019}.

It is intriguing that although Moreton waves are observed in the solar chromosphere, they are not due to magnetoacoustic waves propagating in the chromosphere horizontally, since their propagation speed, 500--2000 km s$^{-1}$, is around 10 times larger than the magnetoacoustic wave speed in the chromosphere. Noticing that their propagation speed is similar to the magnetoacoustic wave speed in the corona, it was proposed that, as a solar flare occurs, the high gas pressure in the flare excites a fast-mode magnetoacoustic shock wave propagating with a speed of 500--2000 km s$^{-1}$ in the solar corona \citep{Uchida1968, Uchida1974}. The footpoint of the coronal shock wave sweeps the solar chromosphere to form Moreton waves as a secondary phenomenon. According to this model, which is called Uchida model, there should be coronal counterparts of chromospheric Moreton waves \citep{Chen2011}, which were later confirmed by extreme ultraviolet (EUV) imaging observations, i.e., those sharp wave fronts in the Solar and Heliospheric Observatory (SOHO) observations \citep{bies02, warmuth04, Thompson09} and the fast-component EUV waves in the Solar Dynamics Observatory (SDO) data \citep{Pesnell_2012, chenwu11, asai12, nitta2013}. The Uchida model works well in principle, but still has been refined further \citep{vrsnak16, Chen2016, zheng23, zhong25}. For example, it has been suggested that the coronal shock wave is driven by a CME, rather than due to high gas pressure of a solar flare \citep{Cliver1999, Chen2002, Chen2005, francile2016, jinm18, zheng2024}. With observations from new telescopes, it is expected that more observational features will be revealed that were missed by the original Uchida model.

In most cases, as expected, Moreton waves decay monotonically as they propagate further away from the flare site \citep{warmuth01, warmuth2015}. However, with the imaging spectroscopic observations of the Chinese H\(\alpha\) Solar Explorer (CHASE), we reveal in this paper that the Moreton wave on 2022 October 2 was locally enhanced midway during propagation. It would be interesting to explore why the Moreton wave becomes locally brighter midway through its propagation.

\section{Observations and numerical method}

\subsection{Observational data}\label{obs}
On 2022 October 2, a filament eruption occurred at 20:19:00 UT, which evolved to a CME and was accompanied by an EUV wave lasting from 20:19 to 20:27 UT.
We observed this event by using the Atmospheric Imaging Assembly \citep[AIA;][]{Lemen_2012} onboard SDO, which observes the full Sun in 7 EUV and 2 UV passbands. The EUV channels have a 12~s cadence and $0\farcs6$~pixel size, providing continuous observations for studying EUV wave kinematics. The Fe~\textsc{ix} 171~\AA\ and Fe~\textsc{xii} 193~\AA\ channels cover plasma emissions from quiet to active corona, corresponding to temperatures of (0.6--1.5)$\times$10$^6$~K. The Helioseismic and Magnetic Imager \citep[HMI;][]{Scheerrer2012} provides full-disk line-of-sight and vector magnetograms with cadences of 45~s and 720~s, respectively, and a pixel size of $1\farcs0$, using the Fe~\textsc{i} 6173~\AA\ photospheric line. To enhance EUV wave signatures, we applied a base-difference technique to reveal multiwavelength temporal evolution of the wave. 
For the magnetic-field extrapolation, we used the pre-eruption HMI vector magnetogram obtained at 20:12:00 UT on 2022 October 2. The data were tracked and cropped to the region of interest, the 180$^\circ$ ambiguity was resolved, and the vector magnetic field was transformed into the heliographic coordinate system. Force-free preprocessing was then performed before the magnetic-field extrapolation.
Since the wave propagated across nearly a quarter of the solar disk (including quiet-Sun regions), we computed the local potential field via the Poisson equation using the poloidal–toroidal decomposition method \citep{Chandrasekhar1961, Backus1986, Fisher2020}. The squashing factor $\log Q$ calculation follows the improved algorithm \citep{Scott2017}, which reduces computational cost by using differences in basis vectors instead of field line differencing to compute the norm. 
The magnetic field lines were integrated using an adaptive fourth-/fifth-order Runge--Kutta method \citep{Dormand1980}. Magnetic null points were subsequently refined with the Newton--Raphson method (see the Appendix for details). After all converged null solutions were obtained, repeated solutions at almost the same location were removed.

\begin{figure*}
	\centering
	\includegraphics[width=0.76\textwidth]{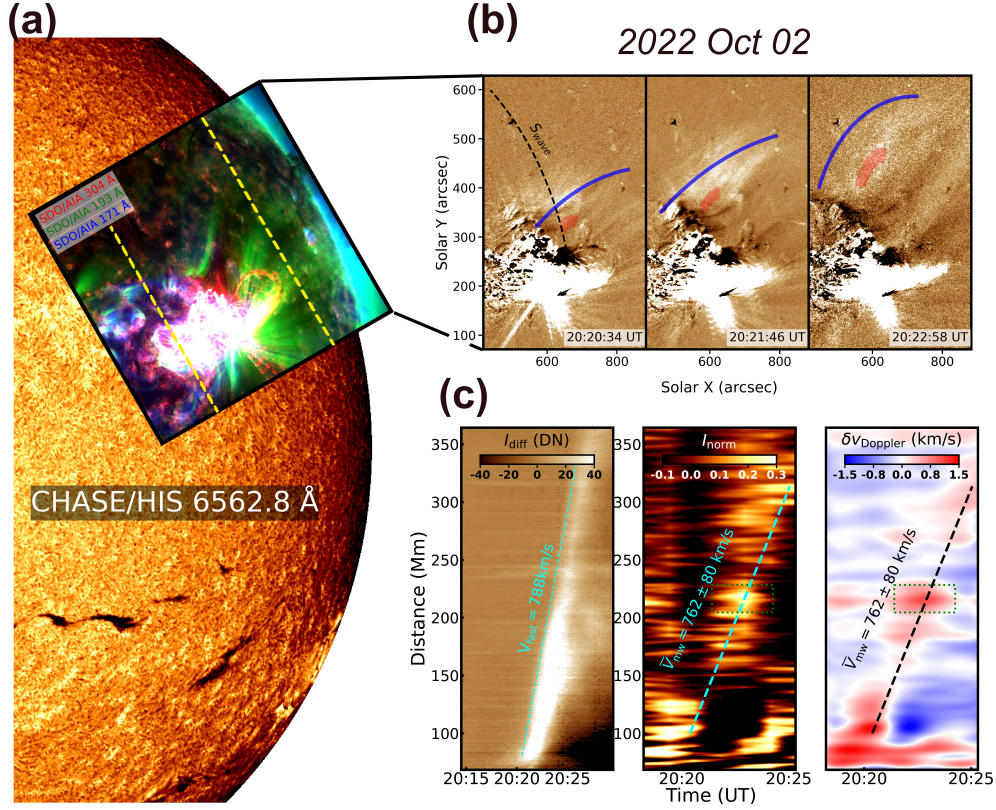} 
	\caption{Multi-wavelength observations of the shock wave driven by the solar eruption on 2022 October 2.
This figure presents the EUV wave, chromospheric Moreton wave associated with the X1.0-class flare and CME event. Panel (a): Full-disk H\(\alpha\) image from CHASE/HIS overlaid with a three-color EUV composite (Red: 304~\AA\, Green: 193~\AA\, Blue: 171~\AA\ ) from SDO/AIA at 20:21:48 UT, showing the active region and the CME. Panel (b): Time evolution of the coronal shock wave. Base-difference images of the \ion{Fe}{12} 193~\AA\ channel from SDO/AIA observations at three consecutive times. Blue solid lines represent the fronts of the EUV wave at different times, and red contours show the H\(\alpha\) Moreton wave front at each time. The black dashed line in the \textit{left panel} indicates the slice \(S_{\mathrm{wave}}\) for the time–distance diagrams. Panel (c): Multi-wavelength time--distance diagrams along the \(S_{\mathrm{wave}}\). Left: 193~\AA\ base-difference intensity. Middle: H\(\alpha\) intensity. Right: H\(\alpha\) Doppler velocity. Both middle and right panels trace the Moreton wave at 762 \(\pm\) 80 km s\(^{-1}\).
See also movie~1.}
	\label{fig1}
\end{figure*}

The Moreton wave was resolvable in H$\alpha$ during 20:19–20:25 UT. The H$\alpha$ Imaging Spectrograph (HIS) on board CHASE obtains full-disk spectral data in the ranges 6559.7--6565.9~\AA\ and 6567.8--6570.6~\AA\ via raster scanning mode (RSM), with a full-disk scan time of 46~s. For the two-binning mode, the spectral resolution is 0.048~\AA\ per pixel, and the pixel size is about $1\farcs04$~in space. The preprocessing pipeline includes dark-field and flat-field correction, slit-image curvature correction, and wavelength calibration \citep{Qiu2022}. To enhance Moreton wave signatures, we computed blue–red wing contrast images as
\[
I(x,y) = \frac{B - R}{B + R},
\]
where $B$ and $R$ are the integrated intensities over $\pm$0.5~\AA\ from the line center in the blue and red wings, respectively, with an integration bandwidth of 0.1~\AA. The full H$\alpha$ spectra allow Doppler velocity inversion via the cross-correlation method. We select an average quiet-Sun spectrum $\overline{I(\lambda)}$ from a $100\times100$ pixel region at disk center as a reference. The convolution of H$\alpha$ spectra is calculated, and a polynomial fit around the maximum is used to precisely determine the Doppler velocity. In addition, the chromospheric Dopplergram is processed after removing large-scale flows including gravitational redshift, differential rotation \citep{Rao2024}, meridional circulation, and limb shift.
The wavelength displacement is converted to the line-of-sight velocity through
$v_{\rm LOS}=c_0\Delta\lambda/\lambda_0$, where $\lambda_0=6562.8$~\AA. We determine the Doppler velocity at each spatial point relative to a reference spectrum averaged over a quiet Sun region. Its uncertainty therefore includes both the uncertainty in this average and the error in fitting the spectral shift.
For the spatial Doppler profiles shown in Figure~\ref{fig2}, we fit a Gaussian enhancement superposed on a linear background,
\begin{equation}
f(s)=as+b+A\exp\left[-\frac{(s-s_0)^2}{2\sigma_s^2}\right].
\end{equation}
The fitting is performed using nonlinear least squares weighted by the profile scatter.
The wave velocities in Figures~\ref{fig1} and~\ref{fig3} are obtained from linear fitting to the bright ridges in the time--distance diagrams. For Figure~\ref{fig4}d, the propagation speed is derived from the local slope of the fitted ridge.

\subsection{Simulation Setup} 
We developed a 3D CME evolution model to investigate the mechanism of local Moreton wave enhancement. Based on the observed photospheric magnetogram, we simulated a solar eruption in a realistic atmosphere to reproduce the mode conversion of the EUV wave associated with a magnetic null point. The simulation domain includes both active regions (with sunspots) and quiet-Sun regions, requiring treatment of strong-field and low-$\beta$ plasma. 
Here $r$ is the heliocentric radius, $\theta$ is the colatitude, and $\phi$ is the longitude, with the origin at Sun center. The eruption source and the wave-propagation direction in the numerical domain were registered through the coalignment of the HMI magnetic field with the SDO/AIA observations.
To handle large Alfvén speeds ($B/\sqrt{\mu_0 \rho}$) while maintaining a reasonable time step, we solved the semi-relativistic MHD equations \citep{Gombosi2002}, including gravity, background heating, field-aligned thermal conduction, radiative loss, and finite resistivity \citep{Guo2025}. We mapped $\boldsymbol{E}$, $\boldsymbol{v}$, and $\boldsymbol{B}$ onto a staggered grid to keep $\nabla\cdot\boldsymbol{B}=0$.

The whole atmosphere was constructed after magnetostatic relaxation. During this relaxation, we set the temperature stratification following a tanh function across the transition region and assumed a constant thermal conduction flux in the corona. The hydrostatic equation was then solved to meet 1D hydrostatic equilibrium. The potential field was incorporated, and the system was relaxed until velocities became negligible (much smaller than the local magnetosonic speed). When magnetic and thermal distributions were stabilized, the magnetostatic relaxation phase was completed. The relaxed state served as the initial condition for the CME simulation. To initiate the eruption, a regularized Biot--Savart law (RBSL) flux rope \citep{Titov2018} was embedded along the polarity inversion line (PIL), mimicking the filament channel geometry with artificial filament threads at the bottom of magnetic dips. 
The details are described in Appendix~\ref{app:setup}.

We adopted the second-order three-step Harten–Lax–van Leer scheme with a Van Leer limiter to solve the MHD equations. The constrained transport method was adopted to preserve $\nabla\cdot\boldsymbol{B}=0$ \citep{Gardiner2005}. Boundary conditions were set as follows: zero velocity at the bottom boundary, outflow at all other boundaries; bottom ghost-cell magnetic fields were extrapolated using a third-order one-sided equal-gradient method, while the zero-gradient extrapolation was used for other boundaries; bottom density and pressure were fixed, top density and pressure were extrapolated hydrostatically, and zero-gradient extrapolation was used for all side boundaries. We used radially uniformly stretched spherical coordinates in the following domain: $r\in[1,\,1.5]\,R_\odot$, $\theta\in[10^\circ,\,85^\circ]$, and $\phi\in[-1^\circ,\,44^\circ]$, with 4-level adaptive mesh refinement achieving an effective resolution of $768\times384\times384$ and a highest resolution of $452$~km $\times 0.2^\circ \times 0.12^\circ$. Note that magnetograms with a higher spatial resolution would result in more magnetic null points, but the characteristic heights and their horizontal distribution remain nearly unchanged.

\subsection{Radiation Synthesis} 
To visualize the EUV wave in simulations, we synthesized EUV emissions based on the density and temperature distributions. Using SDO/AIA response functions, we employed the Radiation Synthesis Tools \citep{Fu2026} to create synthetic images:
\[
I_{\rm AIA}(r, \theta, \phi) = \int G_{\rm AIA}(T)\, n_e^2(r, \theta, \phi) \, ds,
\]
where $G_{\rm AIA}(T)$ is the SDO/AIA response function in the EUV channels, $n_e$ is the electron number density, and $I_{\rm AIA}$ is the integral of radiation along the actual observational line of sight. 
We used the response functions including the CHIANTI correction discussed by \cite{Boerner2014}.
To match observations, we enhanced wave signatures by base-differencing images, using the initial synthetic image as the reference.
For the simulated Moreton wave, given chromospheric spatial resolution limitations, we used the RH code \citep{Rybicki1992, Pereira2015} and combined the plasma velocity field from the simulation with the FAL-C reference atmosphere \citep{Fontenla1993} to compute chromospheric spectral lines, obtaining H$\alpha$ spectra along the EUV wave propagation direction.

\section{Results}

The GOES soft-X-ray emission began to rise at around 20:14:05 UT, and the rapid filament ejection occurred at around 20:19:00 UT. An X1.0-class solar flare and a CME with a speed of 1086 km s\(^{-1}\) occurred. The eruption was simultaneously captured by CHASE/HIS \citep{Li2019, Li2022} and SDO/AIA. Figure \ref{fig1}a presents a full-disk H\(\alpha\) image from CHASE, overlaid with a three-color composite EUV image (Red: 304~\AA\, Green: 193~\AA\, Blue: 171~\AA\ ) from SDO/AIA at 20:21:48 UT, highlighting the active region and the associated eruption. The eruption produced wave perturbations from the solar corona down to the chromosphere.

Figure \ref{fig1}b displays three snapshots of the SDO/AIA 193~\AA\ images from 20:20:34, 20:21:46 and 20:22:58 UT, respectively. A coronal EUV wave is seen to propagate northward, with wave fronts indicated by the blue lines. At the same time, the CHASE H\(\alpha\) observations revealed a chromospheric Moreton wave propagating northward. To illustrate the spatial relationship between the Moreton wave and the EUV wave, we overlay the Moreton wave fronts with red contours on the three subpanels of Figure \ref{fig1}b at the closest instants. It is seen that the chromospheric Moreton wave fronts are nearly cospatial with the coronal EUV wave, lagging behind slightly. To trace the wave kinematics, we select a slice ($S_{wave}$) along the propagation direction as marked in Figure \ref{fig1}b. Figure \ref{fig1}c presents the time--distance diagrams of the 193~\AA\ base-difference intensity (left), H\(\alpha\) intensity (middle), and H\(\alpha\) Doppler velocity (right) along the slice. It is revealed that the EUV wave front propagated at a speed of 788\(\pm\)21 km s\(^{-1}\), whereas the Moreton wave, characterized by the bright ridge in the H\(\alpha\) intensity and the redshift ridge in the Doppler velocity diagrams, moved at 762\(\pm\)80 km s\(^{-1}\) (see also movie 1). It is seen that the Moreton wave front lagged slightly behind the EUV wave front and exhibited a redshift followed by a blueshift, characteristic of chromospheric compression. From Figure \ref{fig1}c, it can be seen that the red-shifted Doppler velocity of the Moreton wave was initially strong around the distance of 100 Mm, which is still close to the flare, and then decayed rapidly as typically observed in previous events. However, a striking feature of Figure \ref{fig1}c is that, when the Moreton wave reached a distance of $\sim$213$\pm$15 Mm, it intensified again. 

\begin{figure*}
	\centering
	\includegraphics[width=0.6\textwidth]{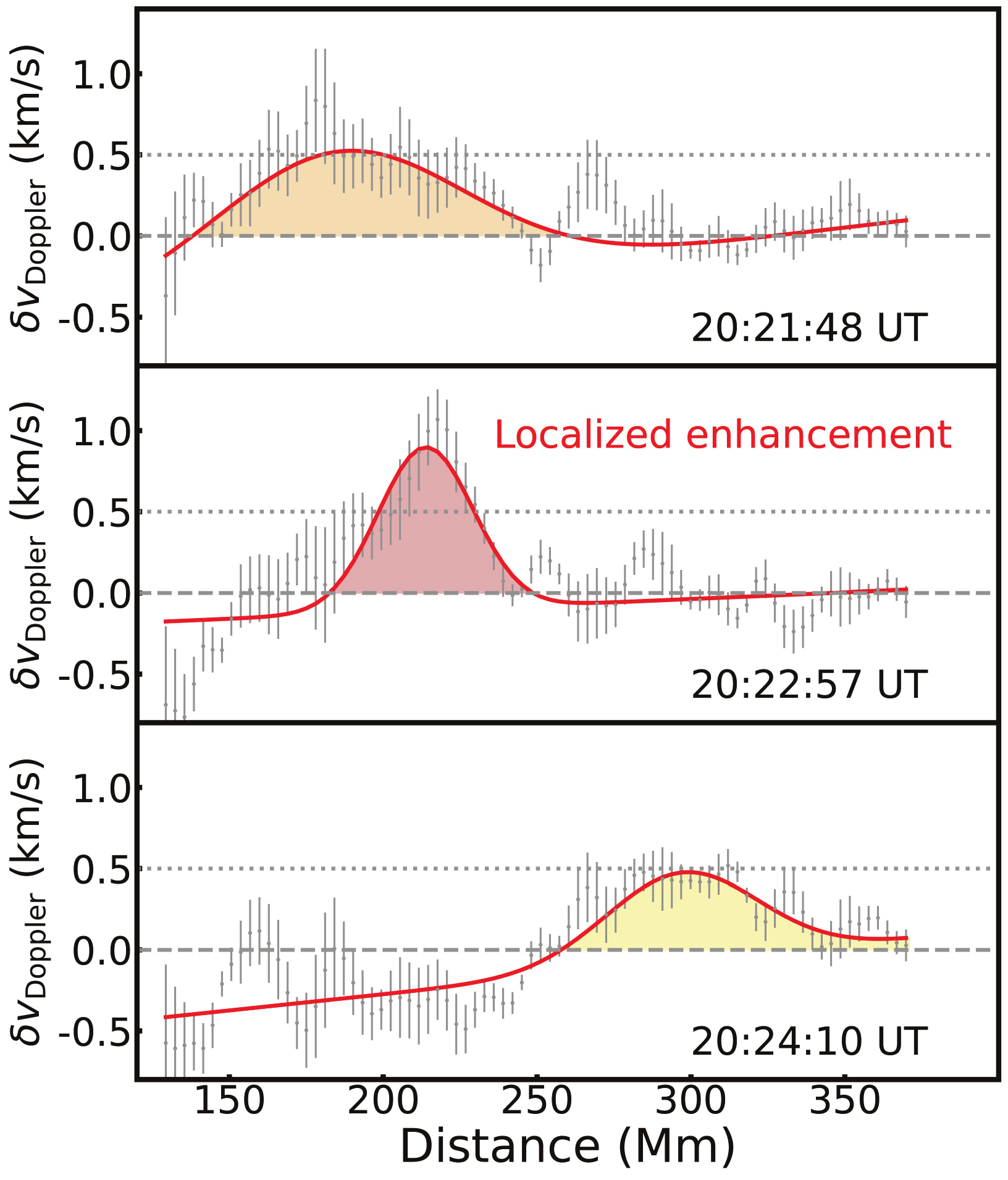} 
	\caption{
Locally enhanced Moreton wave and its Doppler-velocity evolution.
Residual Doppler velocity along the slice $S_{\mathrm{wave}}$ at 20:21:48, 20:22:57, and 20:24:10 UT. Gray points with error bars show the observed Doppler velocities, and red curves show a Gaussian enhancement superposed on a linear background. The Doppler-velocity uncertainties include both the uncertainty of the averaged quiet-Sun reference spectrum and the error in fitting the spectral shift, as described in Section~\ref{obs}. The time-averaged Doppler velocity is $v_{\mathrm{avg}}=3.6$~km~s$^{-1}$.
}
	\label{fig2}
\end{figure*}

To verify this feature, we select three consecutive instants: the pre-enhancement stage (20:21:48 UT), the enhancement stage (20:22:57 UT), and the decay stage (20:24:10 UT). To clarify the local enhancement more clearly, we plot the Doppler velocity distribution along the propagation path at the three instants in Figure \ref{fig2}. For better visualization, we show \(\delta v=v_{\mathrm{Doppler}}-v_{\mathrm{avg}}\), where \(v_{\mathrm{avg}}=3.6\) km s\(^{-1}\) is the time-averaged component of the Doppler velocity. The data points with error bars represent the original observations. It is evident that at 20:22:57 UT, the fitted Doppler velocity reaches its maximum at 0.9 $\pm$ 0.05 km s\(^{-1}\) above the time-averaged component, i.e., $\approx$4.5 km s\(^{-1}\), at a distance of 213.7 $\pm$ 0.8 Mm. This value is $\approx$0.4 km s\(^{-1}\) higher than the fitted peak 69 s earlier.

\section{Discussion}

\subsection{Mode conversion}

\begin{figure}
	\centering
	\includegraphics[width=1.0\textwidth]{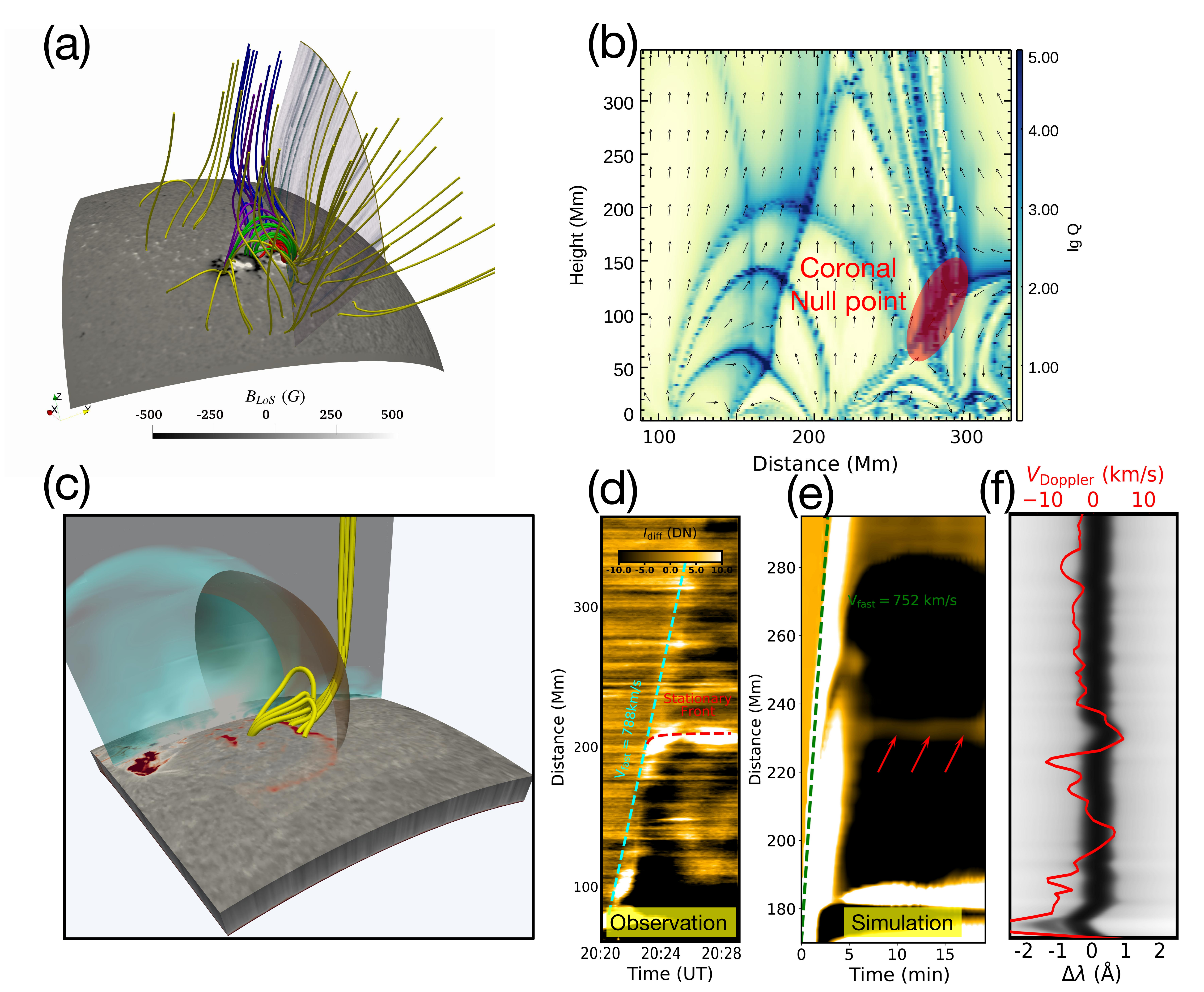} 
	\caption{Panel (a): Potential field extrapolation from HMI vector magnetogram. Panel (b): Squashing factor $\log Q$ map computed using the improved algorithm of \citet{Scott2017}. The black arrow indicates the direction of the vector magnetic field alongside the plane. Panel (c): Synthetic stereoscopic view of the erupting region. The bottom surface represents H\(\alpha\) \(\pm\) (0–1)~\AA\ spectral images. Red contours indicate H\(\alpha\) Doppler redshift velocity, mapping chromospheric compression. Yellow lines are extrapolated magnetic field lines, the orange shell marks the EUV wave front and the side-view plane displays the density compression during the eruption. Panel (d): Observed time--distance diagram of the Fe\,\textsc{ix} 171~\AA\ intensity. Cyan dashed: propagating front (788 km s$^{-1}$); Red dashed: stationary front (initial 151.3 km s$^{-1}$). Panel (e): Time--distance diagram of the synthesized Fe\,\textsc{ix} 171~\AA\ intensity. Green dashed: simulated shock wave front (752 km s$^{-1}$); Red arrows: stationary front. Panel (f): Synthesized H$\alpha$ spectrum at the enhancement stage.
        }
	\label{fig3}
\end{figure}

Since the coronal shock wave is the primary agent for the chromospheric Moreton wave, we examine the coronal EUV wave and magnetic topology around the distance of 210--230 Mm, where the local enhancement of the Moreton wave occurred. It is noticed that the magnetic field always plays a significant role in linking the coronal shock wave to the chromospheric Moreton wave. As illustrated in Figure \ref{fig3}a, we extrapolate the coronal magnetic field from the SDO/HMI magnetogram based on the potential field model. To elaborate the magnetic structure and topology around the Moreton wave enhancement site, we select a cross-section along the wave propagation path, as indicated by the shaded plane in Figure \ref{fig3}a, and then calculate the squashing factor $Q$. The $Q$-map is displayed in Figure \ref{fig3}b as the color scale. It is found that there exists a magnetic null point at around (275, 70) Mm. From the null point extend X-shaped magnetic separatrices, one of which maps downward to the solar surface at the horizontal distance of $\sim$230 Mm, where the local enhancement of the Moreton wave appeared. 

It has been proposed that fast-mode magnetoacoustic waves in the corona can undergo mode conversion in the vicinity of magnetic null points, with the resulting slow-mode waves subsequently traveling along magnetic field lines parallel to a separatrix and eventually concentrating at its footpoint \citep{mclau06, Chen2016, Guo2023}. This theoretical picture naturally prompts us to examine the EUV observations for evidence of such a conversion. For this purpose, we plot the time--distance diagram of the SDO/AIA 171~\AA\ intensity in Figure \ref{fig3}d, which shows that around the distance of 210--230 Mm, a much slower wave with an initial speed of 151.3 km s$^{-1}$ emanated from the coronal fast-mode shock wave. This slow wave subsequently became almost stationary, appearing as a localized stationary front, indicative of mode conversion happening locally. 
The equipartition layer where the Alfvén speed ($v_A$) becomes comparable to the local sound speed ($c_s$), corresponding to $\beta\simeq2/\gamma$, provides a natural region for fast-to-slow magnetosonic mode conversion.
Besides, the conversion is enhanced when the wave vector is nearly aligned with the magnetic field and when the magnetic field varies rapidly across the conversion region.
In most of the solar corona, $\beta$ is much smaller than 1 \citep{rodr19}, meaning the Alfvén speed significantly exceeds the sound speed. Near magnetic null points, however, the diminishing magnetic field reduces the Alfvén speed, thereby enabling the condition for mode conversion to be satisfied. Consequently, the fast-mode coronal shock wave can partially convert into a slow-mode magnetoacoustic wave at such locations. This conversion process naturally explains why the weakening fast-mode coronal EUV wave became locally strengthened at a distance of 210--230 Mm \citep{Chen2016a}. The newly generated bright slow-mode shock then propagates along the local magnetic field lines from the corona down to the chromospheric footpoints, precisely generating the locally enhanced Moreton wave front.

\subsection{Validation via numerical simulation}

In order to reproduce the locally enhanced chromospheric Moreton wave front, we conduct numerical simulations of the shock wave propagation following a solar eruption using the three-dimensional magnetohydrodynamics code MPI-AMRVAC \citep{Xia2018, kepp23}. To obtain the initial conditions for the simulation, the extrapolated potential magnetic field mentioned above is coupled with a hydrostatic equilibrium atmosphere, which is then relaxed to a quasi-static equilibrium. As noted above, a magnetic null point appears near the region where the stationary EUV wave front was observed. The magnetic configuration around this null point consists of large-scale open magnetic field lines above the null and small-scale closed magnetic loops below it. As the initial setup, a magnetic flux rope supporting a solar filament is inserted into the source active region \citep{Guo2025}; the flux rope then erupts due to the upward Lorentz force. The erupting filament drives a rapidly propagating coronal shock wave. 
As shown in Figure \ref{fig3}c, we use the region of strong density compression along the wave-propagation direction to identify the coronal shock front. The shock front forms a dome-shaped structure in the corona, slightly preceding the chromospheric Moreton wave front in the simulation. To facilitate a direct comparison with observations, we synthesize the 171~\AA\ emission based on the simulation results. As the fast-mode EUV wave approaches the magnetic null point, the EUV wave front becomes weakened and leaves a nearly stationary front.

To examine the propagation of the coronal EUV wave from a top view, we take the same slice as defined in Figure \ref{fig1}b. The time--distance diagram of the synthesized 171~\AA\ intensity along this slice is shown in Figure \ref{fig3}e, revealing a coronal EUV wave propagating outward with a nearly constant speed of 752 km s$^{-1}$. As this wave propagates through the corona, it compresses the underlying chromosphere, causing it to move downward. The synthesized H$\alpha$ spectral profile along the same slice at $t$=300 s is plotted in Figure \ref{fig3}f, where a significant redshift at the distance of 230 Mm marks the enhancement of the Moreton wave, consistent with the observed local enhancement shown in Figure \ref{fig2}. These simulation results thus provide strong support for the scenario in which mode conversion around the magnetic separatrix is responsible for both the stationary EUV wave front and the locally enhanced Moreton wave.

\begin{figure}
	\centering
	\includegraphics[width=0.84\textwidth]{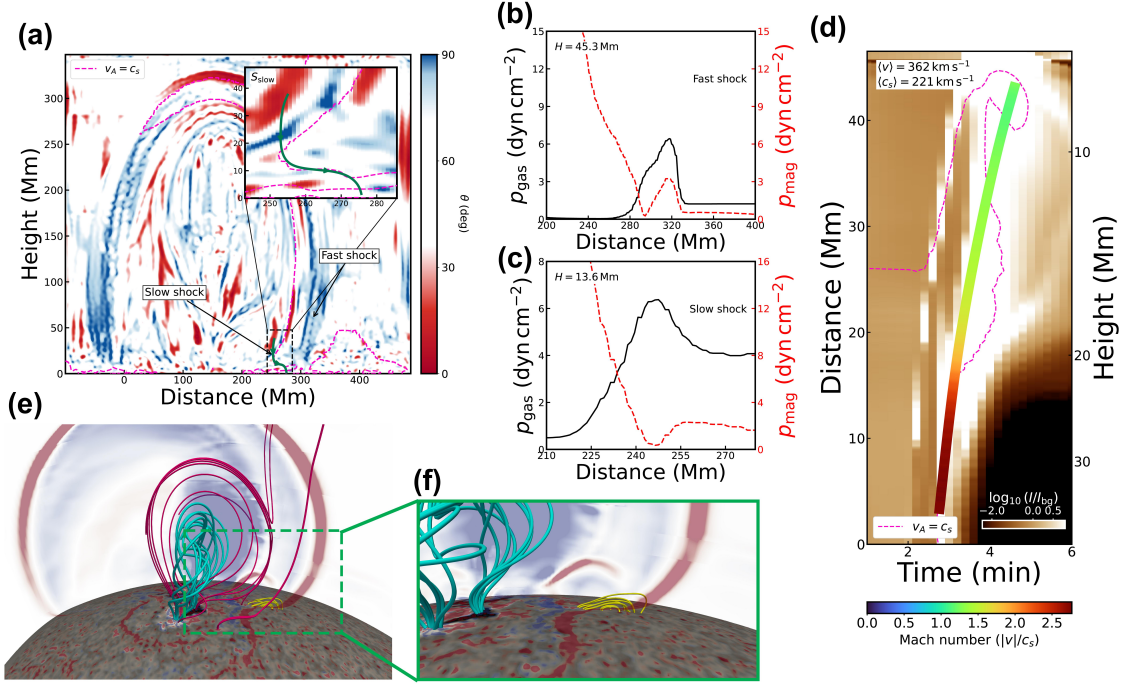} 
	\caption{
Diagnostics of the fast-to-slow magnetosonic mode conversion.
(a) Acute angle between the velocity perturbation $\delta\boldsymbol{v}$ and the magnetic field $\boldsymbol{B}$. The green curve marks the sampling path $S_{\rm slow}$, with an enlarged view shown in the inset.
(b)--(c) Gas pressure (black solid curves) and magnetic pressure (red dashed curves) across the fast- and slow-mode fronts, respectively.
(d) Time--distance diagram of $\log_{10}(I/I_{\rm bg})$ along the slice $S_{\rm slow}$, where $I_{\rm bg}$ is the initial intensity at each position. The slice is a sampling path in the low corona extending downward from a height of $\sim$38~Mm to the chromosphere; the left axis gives the distance along the slice measured from its upper end, and the right axis the corresponding altitude, which decreases downward. The fitted bright ridge is color-coded by the acoustic Mach number. The magenta dashed curves mark $v_A=c_s$.
(e) Side view of the magnetic configuration beneath the coronal null point.
(f) Enlarged view of the magnetic structure and stationary front above the locally enhanced Moreton-wave region. Redshift and blueshift denote plasma motions away from and toward the observer, respectively. See also movie~2.
}
\label{fig4}
\end{figure}

As the shock wave sweeps the magnetic loops around the null point, the wave front crosses the equipartition layer where $v_A\simeq c_s$, causing part of the fast-mode magnetoacoustic wave to convert into a slow-mode shock wave. Figure~\ref{fig4}a shows the angle between the velocity perturbation and the magnetic field: The red regions indicate that the perturbation is nearly parallel to the magnetic field.
As illustrated in Figures \ref{fig4}b and \ref{fig4}c, the variations in gas pressure and magnetic pressure differ distinctly between the two wave modes. For the fast-mode wave, both gas pressure and magnetic pressure increase simultaneously across the wave front. By contrast, the slow-mode shock occurs where the magnetic pressure decreases while the gas pressure increases. 
To verify the converted slow-mode shock, we construct a time--distance diagram along the slice $S_{\rm slow}$, as shown in Figure~\ref{fig4}d. The brightest ridge traces its downward propagation as indicated by a thick line, which is color-coded by $M_{\rm prop}=|{\rm d}s/{\rm d}t|/c_s$, where $s$ is the distance along the slice and $c_s$ is calculated from the mean temperature of each segment. The averaged propagation speed and sound speed are 362 and 221~km~s$^{-1}$, respectively. The magenta dashed curves mark the $v_A=c_s$ layer.
From the side view of the eruption (Figure \ref{fig4}e), it is evident that when the CME piston-driven coronal shock passes through the horizontal magnetic loops beneath the null point, it generates a slow-mode shock. Because the slow-mode shock propagates along the magnetic field lines \citep{Bogdan2003}, it forms a stationary front from the top view, separated from the weakened fast-mode wave that continues to move outward (see also movie 2). A zoom-in view of the null-point region proximity (Figure \ref{fig4}f) reveals that the stationary front is nearly parallel to the magnetic loops and is located directly above the locally enhanced Moreton wave. As the converted disturbance propagates downward, the acoustic Mach number $M_{\rm prop}$ decreases. When the slow-mode shock wave reaches the chromosphere, a strong downward movement is generated, producing an enhanced H$\alpha$ redshift. It is noted that the recorded time of H$\alpha$ raster in the CHASE data is at the beginning of the scanning, therefore, the actual time of the Moreton wave should be tens of seconds later in each raster.

To test whether the coronal null point is crucial for the stationary EUV wave front and hence the enhanced Moreton wave, we also perform simulations in a control case where the photospheric magnetogram below the null point is modified so that there is no null point in the extrapolated coronal magnetic field. No stationary front is found in the control run, as shown in Appendix~\ref{app:control}.

\subsection{Possible ubiquity of mode conversion}

After stationary EUV wave fronts were discovered in the wake of fast-mode waves \citep{chandra2016}, they were explained as being due to mode conversion from a fast-mode wave to a slow-mode wave \citep{Chen2016, Chandra2018, Chandra2022}. Later, this mode conversion was confirmed to be universal in EUV wave events \citep{chandra2016, chan21, kumar24, razquin26}. With the 2022 October 2 event, we further found that the bright slow-mode EUV wave can propagate downward from the corona to the chromosphere, forming a locally enhanced Moreton wave front. By analyzing the multiwavelength observational data and conducting MHD numerical simulations, we propose that such an unexpected enhancement is due to wave mode conversion, i.e., the initial fast-mode MHD shock wave driven by the eruption is efficiently converted to a slow-mode MHD wave, which is trapped in the local magnetic field lines, hence forming the local enhancement of the Moreton wave.

\begin{figure}
	\centering
	\includegraphics[width=0.84\textwidth]{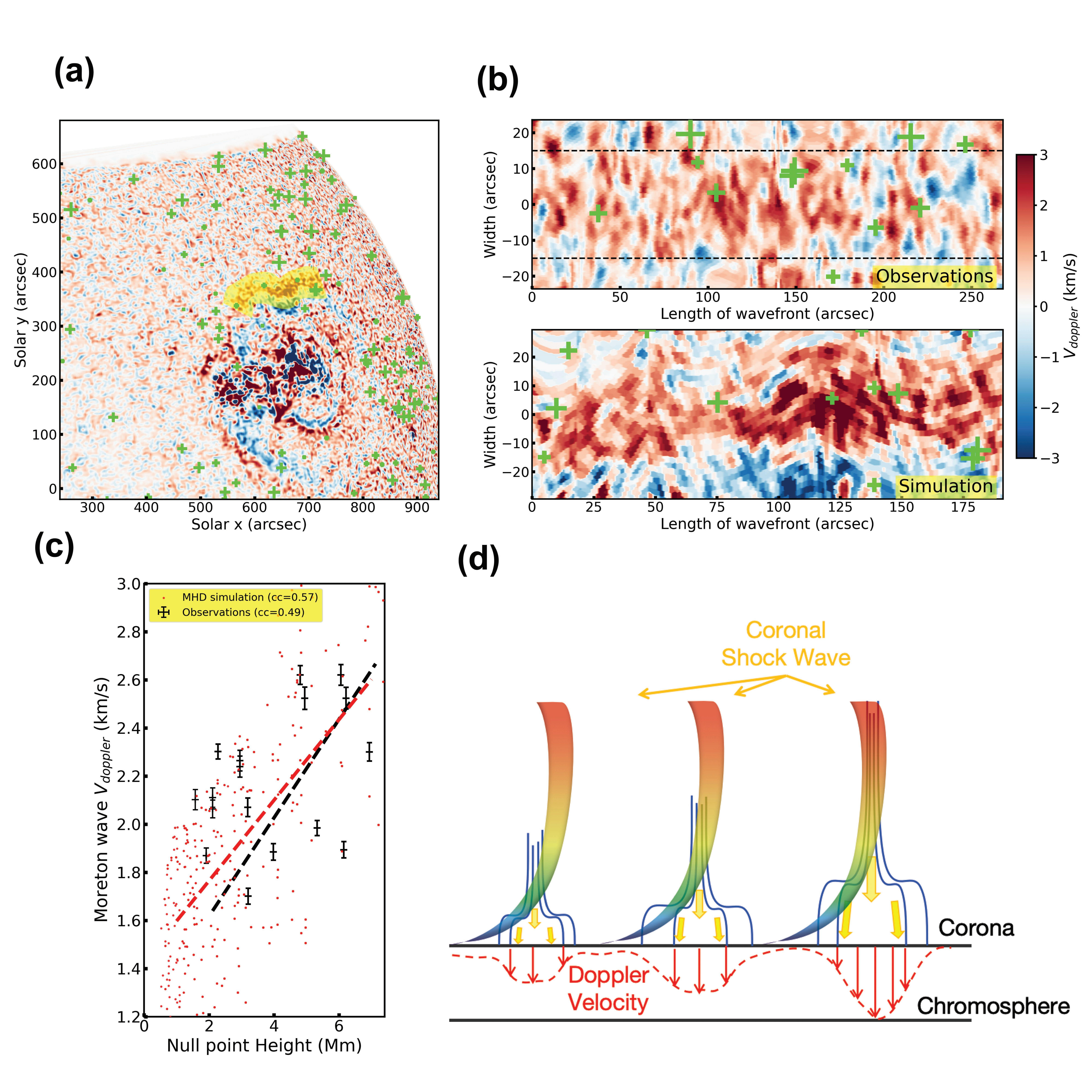} 
	\caption{Relationship between Moreton wave front Doppler velocities and coronal magnetic null points. Panel (a): Dopplergram from MHD simulation with coronal magnetic null points (plus signs) overlaid. The yellow contour illustrates the Moreton wave front at this moment.  
    Panel (b): Zoomed-in distributions of the H$\alpha$ Dopplergram and magnetic null points. The color scale represents the Doppler velocity, and the size of each plus sign is determined by the altitude of the null point.  
    Panel (c): Doppler velocity at the wave front versus height of coronal magnetic null points. Red dots represent MHD simulation results, while black dots with error bars show observational data. 
    Both red and black dashed fitted lines indicate a linear correlation between Doppler velocity and null point height.  
    Panel (d): Schematic sketch illustrating how higher null points produce stronger compression.
        }
	\label{fig5}
\end{figure}

Since there are a plethora of magnetic null points in the low corona immediately above the solar surface \citep{regn08}, it is expected that similar mode conversion would occur more frequently in the low corona. Inspired by the analysis of this event, we speculate that the appearance of Moreton wave fronts in general is facilitated by the existence of magnetic null points in the low corona. To verify this hypothesis, we plot in Figure \ref{fig5}a the H$\alpha$ Dopplergram of our simulation 120 seconds after the filament eruption, where the Moreton wave corresponds to the enhanced redshifts as highlighted by the yellow shade. Similar to observations, the redshifts are immediately followed by blueshifts. All magnetic null points in the low corona are marked by the plus signs. Indeed it is seen that magnetic null points are widespread across the solar surface, and are concentrated around the Moreton wave front. Figure \ref{fig5}b displays the zoomed-in distributions of the H$\alpha$ Dopplergram and magnetic null points around the Moreton wave front for both simulations (bottom subpanels) and observations (top subpanels), where the color scale represents the Doppler velocity, and the size of each plus sign indicates the altitude of the null point. This panel gives the impression that the Moreton wave front has larger Doppler velocity around higher magnetic null points in both simulation and observation. In order to check the relationship between the two quantities, we plot the scatter points of the H$\alpha$ Doppler velocity and the null point height in our MHD simulation as the red dots in Figure \ref{fig5}c. Note that the displayed Doppler velocity is normalized by the averaged Doppler velocity at each time, the uncertainty of the null heights is derived from the deviation of null-point heights from the averaged value within 10 pixels of the wave front in the H$\alpha$ Dopplergram, and the velocity uncertainty is described in Section~\ref{obs}. It is seen that there is a clear positive correlation between the H$\alpha$ Doppler velocity and the null point height in our simulation, which can be fitted with a linear function as indicated by the dashed red line. The data in the observation are overlaid as black points with error bars, which are fitted with the black dashed line. It is found that both simulation and observation reveal a linear correlation between the H$\alpha$ Doppler velocity and the null point height.

The linear relationship between the H$\alpha$ Doppler velocity and the magnetic null point height can be understood as follows: When a filament erupts, it drives a bow shock above the ejecta. The shock is the strongest around the nose part of the shock front, and becomes weaker at lower heights. When the shock wave impinges a magnetic null point at a higher altitude, the incoming fast-mode wave is stronger, hence the converted slow-mode shock is stronger. When the slow-mode shock travels along the magnetic field line down to the chromosphere, the H$\alpha$ redshift would be stronger, i.e., the Moreton wave front would be stronger. It is noted in passing that the widespread magnetic null points indicated in Figure \ref{fig5}a are concentrated along magnetic networks, which might help explain why Moreton wave fronts appear preferentially along supergranular boundaries \citep{zhai26}. The ubiquitous existence of magnetic null points in the solar low corona implies that mode conversion from fast-mode to slow-mode magnetoacoustic waves is a common phenomenon. This process not only generates locally enhanced Moreton waves but also facilitates the dissipation of fast-mode magnetoacoustic wave energy. This redistribution and dissipation of wave energy may also contribute to coronal heating   \citep{mclau06}, although a quantitative heating rate or energy budget is beyond the scope of the present study. 

\section{Summary and Conclusion}

In this work, we analyzed multiwavelength imaging and spectroscopic observations of the Moreton wave and EUV wave event on 2022 October 2. The CHASE observations revealed that the H$\alpha$ Doppler redshift of the Moreton wave first decreased during its propagation and then showed an unexpected local enhancement at a distance of $\sim$213$\pm 15$~Mm. At nearly the same location, the SDO/AIA observations show that the propagating fast-mode EUV wave left behind a much slower, nearly stationary front. Such a coincidence strongly indicates that the local enhancement of the Moreton wave is closely related to the evolution of the coronal EUV wave. The magnetic-field extrapolation showed a coronal magnetic null point at a height of approximately 70~Mm around the enhanced Moreton wave front, with one of its separatrices extending downward toward the location of the Moreton-wave enhancement. Approaching the null point, the Alfv\'en speed decreases and becomes comparable to the local sound speed, providing favorable conditions for fast-to-slow magnetosonic mode conversion. With three-dimensional MHD simulations, we reproduced the propagating fast-mode shock, the nearly stationary front, and the locally enhanced H$\alpha$ redshift. The nearly field-aligned velocity perturbation and the opposite variations between gas and magnetic pressures are consistent with a converted slow-mode shock wave. Its downward propagation exceeds the local sound speed, with the acoustic Mach number $M_{\rm prop}$ decreasing gradually as the slow-mode shock wave approaches the chromosphere. These results support that the local enhancement of the Moreton wave is related to fast-to-slow mode conversion around the magnetic null point. It is noted in passing that, even for an ordinary Moreton wave front, we found a positive relationship between the H$\alpha$ Doppler redshift velocity and the heights of low-lying widespread magnetic null points. Such a relationship suggests that mode conversion may occur more commonly during the propagation of Moreton waves, which becomes an additional factor influencing the intensity of Moreton wave fronts in addition to the strength of the incident shock wave and the local magnetic geometry.

\begin{acknowledgments}
The research was financially supported by NSFC (12127901, 12333009), the National Key R\&D Program of China (2020YFC2201200), and the Fundamental Research Funds for the Central Universities (KG202506). CHASE mission is supported by China National Space Administration. The SDO data are available courtesy of NASA/SDO teams. The numerical calculations in this paper were performed in the cluster system of the High Performance Computing Center (HPCC) of Nanjing University.
\end{acknowledgments}


\clearpage
\appendix

\section{Numerical Setup}
\label{app:setup}

This appendix summarizes the initial magnetic and atmospheric models, thermodynamics, magnetic diffusivity, the parameters used in the semi-relativistic treatment, and the Newton--Raphson method.

\subsection{Initial Magnetic Field}

The initial magnetic field is obtained from the pre-eruption HMI vector magnetogram at 20:12:00~UT on 2022 October 2. The potential field is coupled with a hydrostatic atmosphere and relaxed before insertion of the regularized Biot--Savart law (RBSL) flux rope.

Because the pre-eruption filament is too faint to be detected, the RBSL axis is prescribed following the observed morphology of the filament in the early phase of eruption, which is along the PIL between the two flare ribbons. The observed clockwise rotation of the erupting filament and the right skew of the flare loops both indicate positive helicity of the filament channel. Once the axis path is specified, the flux rope in our implementation is determined by the helicity sign, minor radius, and prescribed magnetic flux. We adopt
\begin{equation}
a = 2.5~{\rm Mm}, \qquad \Phi = 1.0\times10^{21}~{\rm Mx}.
\end{equation}
No additional parameters are independently tuned in constructing the RBSL flux rope. The minor radius and magnetic flux are selected to reproduce the observed large-scale eruption while avoiding excessive initial reconnection between the inserted flux rope and the surrounding magnetic field.

\subsection{Atmospheric and Thermodynamic Prescriptions}

The plasma is assumed to consist of fully ionized hydrogen and helium with a number ratio of 10:1, giving
\begin{equation}
\rho = 1.4m_p n_H, \qquad p = 2.3n_H k_B T, \qquad n_e = 1.2n_H .
\end{equation}
The corresponding mean molecular weight is
\begin{equation}
\mu = \frac{1.4}{2.3},
\end{equation}
and the sound speed is evaluated as
\begin{equation}
c_s = \sqrt{\frac{\gamma p}{\rho}},  \qquad \mathrm{ where }~ \gamma = \frac{5}{3}.
\end{equation}

Using $h=r-R_\odot$ as the height above the solar surface, the initial temperature distribution is prescribed as
\begin{equation}
T(h)=
\begin{cases}
T_{\rm ch}, & h<0, \\[5pt]
T_{\rm ch}+\dfrac{T_{\rm co}-T_{\rm ch}}{2}
\left[1+\tanh\left(\dfrac{h-h_{\rm tr}+f_{\rm tr}}{w}\right)\right], & 0\leq h<h_{\rm tr}, \\[10pt]
\left[\dfrac{7F_c}{2\kappa_c}(h-h_{\rm tr})+T_{\rm tr}^{7/2}\right]^{2/7}, & h\geq h_{\rm tr},
\end{cases}
\end{equation}
where
\begin{equation}
f_{\rm tr}=w\,\operatorname{arctanh}\left[2\frac{T_{\rm tr}-T_{\rm ch}}{T_{\rm co}-T_{\rm ch}}-1\right].
\end{equation}

The adopted parameters are
\begin{equation}
T_{\rm ch}=8000~{\rm K}, \qquad T_{\rm co}=5\times10^6~{\rm K}, \qquad T_{\rm tr}=10^6~{\rm K},
\end{equation}
\begin{equation}
h_{\rm tr}=2~{\rm Mm}, \qquad w=250~{\rm km},
\end{equation}
and
\begin{equation}
F_c=2\times10^5~{\rm erg\,cm^{-2}\,s^{-1}}, \qquad
\kappa_c=8\times10^{-7}~{\rm erg\,cm^{-1}\,s^{-1}\,K^{-7/2}}.
\end{equation}

The initial one-dimensional atmosphere satisfies hydrostatic equilibrium. The bottom pressure is
\begin{equation}
p_{\rm bot}=276.5~{\rm erg\,cm^{-3}},
\end{equation}
and the gravitational acceleration is
\begin{equation}
\boldsymbol g=-g_\odot\left(\frac{R_\odot}{r}\right)^2\hat{\boldsymbol r}, \qquad g_\odot=274~{\rm m\,s^{-2}}.
\end{equation}

The internal-energy equation includes compressional work, field-aligned thermal conduction, optically thin radiative losses, background heating, and Joule heating, i.e., 
\begin{equation}
\frac{\partial e}{\partial t}+\nabla\cdot(e\boldsymbol v)
=-p\nabla\cdot\boldsymbol v
+\nabla\cdot(\boldsymbol\kappa\cdot\nabla T)
-n_e n_H\Lambda(T)
+H_{\rm bgr}
+Q_J,
\end{equation}
where
\begin{equation}
e=\frac{p}{\gamma-1}.
\end{equation}

Thermal conduction is field aligned,
\begin{equation}
\boldsymbol\kappa=\kappa_\parallel\hat{\boldsymbol b}\hat{\boldsymbol b}, \qquad
\hat{\boldsymbol b}=\frac{\boldsymbol B}{|\boldsymbol B|},
\end{equation}
with
\begin{equation}
\kappa_\parallel=10^{-6}T^{5/2}~{\rm erg\,cm^{-1}\,s^{-1}\,K^{-1}},
\end{equation}
where $T$ is expressed in Kelvin.

The optically thin radiative-loss function $\Lambda(T)$ is adopted from \citet{Colgan2008}. The background coronal heating is prescribed as
\begin{equation}
H_{\rm bgr}=H_0\exp\left(-\frac{h}{\lambda_{\rm bgr}}\right),
\end{equation}
with
\begin{equation}
H_0=10^{-4}~{\rm erg\,cm^{-3}\,s^{-1}}, \qquad \lambda_{\rm bgr}=60~{\rm Mm}.
\end{equation}

\subsection{Magnetic Diffusivity}

A localized magnetic diffusivity is applied below the erupting filament, which is described as 
\begin{equation}
\eta_m(\boldsymbol x)=
\begin{cases}
\displaystyle
\eta_b\left[1+\frac{10\zeta}{1+\exp[-(\zeta-0.1)/0.01]}\right], &
\boldsymbol x\in Z^*,\quad |\boldsymbol B|>0, \\[10pt]
0, & \mathrm{otherwise},
\end{cases}
\end{equation}
where
\begin{equation}
\eta_b=500~{\rm km^2\,s^{-1}},
\end{equation}
and
\begin{equation}
\zeta=\left(\frac{|\nabla\times\boldsymbol B|\delta}{|\boldsymbol B|}\right)^2,
\end{equation}
with $\delta$ denoting the minimum local grid spacing.

Here $Z^*$ denotes the prescribed low-lying mask below the erupting filament and below the height of 0.01$R_\odot$. Both the base and current-dependent diffusivity terms vanish outside this region, and the diffusivity is also set to zero where $|\boldsymbol B|=0$.

The quantity $\eta_m$ is the magnetic diffusivity rather than the electrical resistivity. With
\begin{equation}
\boldsymbol J=\frac{\nabla\times\boldsymbol B}{\mu_0},
\end{equation}
the corresponding resistive contribution to the induction equation is
\begin{equation}
-\nabla\times\left(\eta_m\nabla\times\boldsymbol B\right),
\end{equation}
and the associated Joule-heating term is
\begin{equation}
Q_J=\mu_0\eta_m J^2.
\end{equation}

\subsection{Semi-relativistic Treatment}

The semi-relativistic treatment follows \citet{Gombosi2002}. We adopt a constant reduced light speed
\begin{equation}
c_r=6\times10^8~{\rm cm\,s^{-1}}=6000~{\rm km\,s^{-1}}.
\end{equation}
This value is substantially larger than the propagation speeds of the wave fronts analyzed in this study and is introduced only to limit the numerically large Alfv\'en speed in the strong-field regions of the computational domain.

\subsection{Newton--Raphson Method}
To calculate the locations of the magnetic null points in the three-dimensional magnetic field, we select candidate grid cells with a nonzero Poincar\'e index and use their centers as the initial guesses for Newton--Raphson iterations. We interpolate the magnetic field trilinearly and accept a solution when $|\boldsymbol{B}|<10^{-5}$~G. When multiple solutions converge within one grid cell, i.e., 453~km in the radial direction and 1.4--2.4~Mm along the angular directions, they are considered as duplicate detections, and only one solution is kept.

\section{Control Case without the Coronal Null}
\label{app:control}

In the control case, the photospheric magnetogram below the coronal null is clipped to $[-10,~10]$~G in a region of about $30\arcsec\times60\arcsec$ before coronal magnetic field extrapolation. This region spans $1.8^\circ$ in heliographic longitude and $3.6^\circ$ in latitude, as marked by the red rectangle in Figures~\ref{fig:control}(a, b). All other numerical settings remain unchanged. Neither the coronal null point nor the closed loops below it are sustained in the control case. Figure~\ref{fig:control} compares the photospheric magnetograms, the coronal magnetic configurations, and the synthesized 171~\AA\ time--distance diagrams along the same propagation direction between the orthodox and the control cases. No stationary front forms in the control case, whereas it is clearly present in the orthodox case, which supports our interpretation that the stationary EUV wave front and the associated local Moreton wave enhancement result from the wave mode conversion around the magnetic null point.

Consequently, no locally enhanced downward motion is present in the chromosphere at the distance of 215 Mm. Hence, the synthesized H$\alpha$ Moreton wave does not exhibit local enhancement in the control case.

\setcounter{figure}{0}
\renewcommand{\thefigure}{B\arabic{figure}}
\begin{figure}
	\centering
	\includegraphics[width=\textwidth,height=0.78\textheight,keepaspectratio]{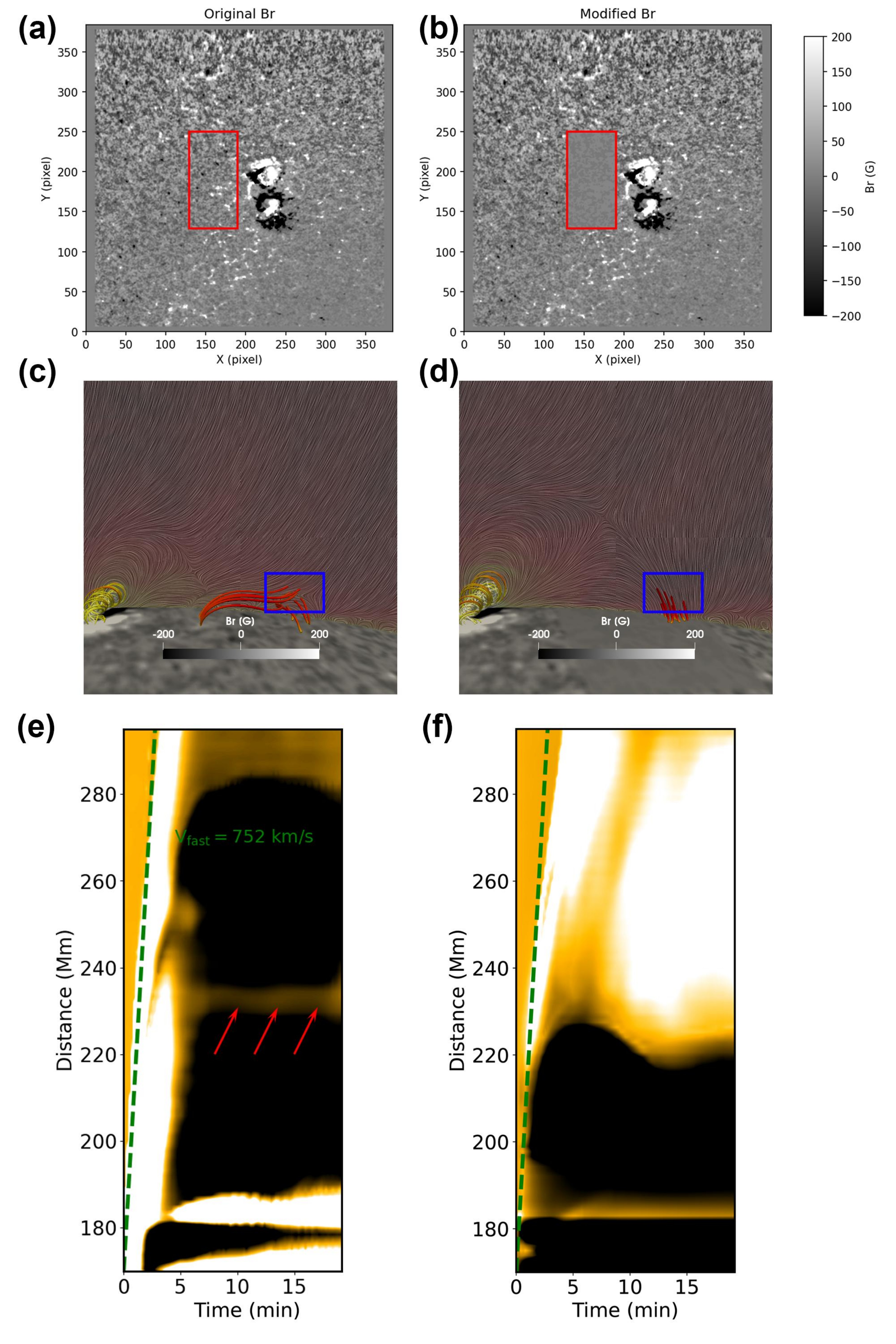}
	\caption{Comparison between the orthodox case (left column) and the control case (right column) without the coronal null point. Panels (a, b): Photospheric magnetograms. In the control case, $B_r$ is clipped over a region marked by the red rectangle. Panels (c, d): Coronal magnetic configurations. Panels (e, f): Synthetic 171~\AA\ time--distance diagrams along the same propagation direction.}
	\label{fig:control}
\end{figure}

\bibliography{ref}{}
\bibliographystyle{aasjournalv7}

\end{document}